 \documentclass[12pt,preprint]{aastex}

\usepackage {epsf}

\newcommand \m {M$_\odot$}
\newcommand \mm {M_\odot}

\begin{document}

\title{Stellar Mass Black Holes in Young Galaxies}

\author{J. Craig Wheeler, Vincent Johnson}
\affil{Department of Astronomy, University of Texas, Austin, TX 78712}
\email{wheel@astro.as.utexas.edu, flint88@mail.utexas.edu}

\begin{abstract}
We explore the potential cumulative energy production of stellar mass
black holes in early galaxies. Stellar mass black holes may accrete 
substantially from the higher density interstellar media of primordial 
galaxies, and their energy release would be distributed more uniformly 
over the galaxy, perhaps providing a different mode of energy feedback 
into young galaxies than central supermassive black holes. We 
construct a model for the production and growth of stellar mass 
black holes over the first few gigayears of a young galaxy. With
the simplifying assumption of a constant density of the ISM,
$n \sim 10^4 - 10^5$ cm$^{-3}$, we estimate the number of accreting stellar
mass black holes to be $\sim 10^6$ and the potential energy 
production to be as high as $10^{61}$ergs over several billion years.
For densities less than $10^5$ cm$^{-3}$, stellar mass black holes
are unlikely to reach their Eddington limit luminosities. The 
framework we present could be incorporated in numerical simulations
to compute the feedback from stellar-mass black holes with inhomogeneous,
evolving interstellar media.


\end{abstract}

\section{Introduction}

Most research on the early formation and growth of black 
holes has focused on the important issues of the growth and 
feedback of intermediate and supermassive black holes; relatively
little attention has been paid to the effects of stellar-mass black
holes and their potential energy production. Estimates of the number
of stellar mass black holes presently in the Galaxy are as high as 
$10^{8}$ to $10^{9}$ \citep{AK02,CDK03}, suggesting that the nearest 
is only parsecs away, but accreting feebly from the modern interstellar 
medium.  The cumulative energy production of $10^{8}$ $10M_{\odot}$ 
black holes could rival that of a single central $10^{9}M_{\odot}$ 
black hole. 

The first stars that formed were extremely low metallicity and 
arguably high mass \citep{BCL02} Population III stars.  
These stars may have only produced intermediate-mass or supermassive 
black holes, if any, but turbulence may broaden the initial
mass function so that lower mass stars and black holes form even
in the first generation \citep{Clark11,Prieto11}. 
The next generation of stars, Population II, were
slightly higher in metallicity and would form stellar mass black holes.
This epoch might have begun at redshift in excess of 15 \citep{BH06}
during the era of reionization, when the interstellar medium number
density, $n_{ism}$, might still be large. In the early formation
of these galaxies, their interstellar medium (ISM) density was probably
much higher than it is currently, perhaps $n_{ism} \sim 10^{4}$cm$^{-3}$
\citep{Greif08}. With ISM densities this high, stellar-mass black holes
could potentially accrete substantial amounts of matter over the course
of the early evolution of galaxies and emit energy back into the 
galaxy and intergalactic medium. Since the number of stellar mass 
black holes could be as high as $10^{8}$, their cumulative energy 
production could be extremely high, given a favorable accretion 
and feedback rate.  Since stellar mass black holes would be 
distributed more evenly over the galaxy than a single supermassive 
black hole, the energy feedback may impact the evolution of a
galaxy in different ways than would a singular source. Stellar-mass
black holes also deliver feedback in a different mode than supernovae,
the energy of which is primarily kinetic, and, unlike supernovae,
black holes persist. The rate of star formation in galaxies is about 
a factor of 100 less than provided by basic estimates \citep{K98,S08}. 
This suggests a feedback process that has not yet been identified. The 
process we outline here may contribute to this regulation of star 
formation in early galaxy evolution.

We apply the prescriptions of black hole accretion with feedback
from \citep{MBCO09} and \citet{PR11} to the stellar mass black hole case to 
approximate the energy production in early galaxies. With these 
estimations, we find that the potential energy production of the stellar 
mass case could be on the order of $10^{60}$ ergs over $\sim 2$ Gyr. 
Section 2 gives an outline of our model for the birth and growth
by accretion of stellar mass black holes. Section 3 addresses the 
rate of accretion of stellar mass black holes including the effects
of radiative feedback and gives estimates
of the number of black holes, their luminosity, and their cumulative
energy liberated as a function of time. Section 4 discusses the possible
constraints of the Eddington limit and \S5 gives our conclusions. 

\section{Star Formation Rate and the Black Hole Birth Rate}

The total number of stars born per unit time per unit galactic mass 
at time, $t$, is given by: 
\begin{equation}
r_{\star}(t)=\int_{0}^{\infty}\dot{n_{\star}}(M_{MS},t)
\Phi(M_{MS})dM_{MS},
\end{equation} 
where $\dot{n_{\star}}(M_{MS},t)$ is the rate of birth of stars 
per unit galactic mass at a given time with main sequence mass, $M_{MS}$, 
and $\Phi(M_{MS})$ is the mass function of main sequence stars with 
mass between $M_{MS}$ and $M_{MS}+dM_{MS}$. The observable star 
formation rate per unit galaxy mass, $s_{\star}$, at a given time is 
then given by:
\begin{equation}
s_{\star}(t)=\int_{0}^{\infty}\dot{n_{\star}} M_{MS} \Phi(M_{MS})dM_{MS},
\end{equation}
where $s_{\star}(t)$ has units of solar masses per year per unit
galaxy mass for $M_{MS}$ measured in solar masses.

Assigning a probability, $P_{BH}(M_{MS})$, that a star of main sequence 
mass, $M_{MS}$, forms a black hole, the black hole birth rate, 
$r_{BH}(t)$, and mass formation rate, $s_{BH}(t),$ per unit galaxy 
mass are given by:
\begin{equation}
r_{BH}(t)=\int_{0}^{\infty}\dot{n_{\star}}\Phi(M_{MS})P_{BH}(M_{MS})dM_{MS},
\end{equation}
and 
\begin{equation}
s_{BH}(t)=\int_{0}^{\infty}\dot{n_{\star}}\Phi(M_{MS})M_{BH,0}(M_{MS})
   P_{BH}(M_{MS})dM_{MS},
\end{equation}
where $M_{BH,0}(M_{MS},t,Z)$ is the initial mass of the black hole
that can, in principle, be a function of the main sequence mass of
its progenitor, the epoch it was born, the metallicity of the progenitor
star and perhaps other parameters. Black holes in binary systems have
measured masses in the range 6 - 20 \m\ \citep{C06,SSMK11}.  
For simplicity, we will scale our results with the assumption
that all black holes are born with a single mass, $M_{BH,0} = 10$ \m.
We can then define the efficiency, $\epsilon_{BH}$, to make black holes from a 
generation of stars at time $t$ as:
\begin{equation}
\epsilon_{BH}\equiv\frac{s_{BH}(t)}{s_{\star}(t)}\backsimeq 
   M_{BH,0}\frac{r_{BH}(t)}{s_{\star}(t)},
\end{equation}
and thus,
\begin{equation}
r_{BH}(t)\simeq\frac{s_{\star}(t)}{M_{BH,0}}\epsilon_{BH}(t).
\end{equation}
The parameter $\epsilon_{BH}$ is proportional to
the fraction of all stars that are O-type stars, $f_{O\star}$, and
the fraction of O-type stars that eventually turn into black holes,
$f_{BH}$. This efficiency could vary with epoch and other parameters, 
but, again for simplicity and illustration we will take it to be constant. 
We adopt $\epsilon_{BH}\sim f_{O\star}f_{BH}\sim 10^{-3}$. This corresponds
approximately to making black holes from stars with $M_{MS} > 30$\m 
(that is, $P_{BH}(M_{MS}$) being a step function at 30 \m) for a 
Salpeter mass function with minimum mass of $\sim 0.1$ \m.
We thus find for the rate of birth of black holes per unit galaxy mass,
\begin{equation}
\label{BHrate0}
r_{BH}(t) \simeq 1\times10^{-4} s_{\star}(t)\epsilon_{BH,-3}M_{BH,0,10}^{-1} 
   ~yr^{-1}
\end{equation}
where $M_{BH,0,10}$ is the mass of the initial black holes in units of 
10 \m\, $s_{\star}(t)$ is in $M_{\odot}yr^{-1}$ per unit galaxy mass
and $\epsilon_{BH,-3}$ is the black hole production efficiency in 
units of $10^{-3}$. Below we will want to address the rate of star 
formation and the associated rate of production of black holes
over an entire galaxy. We define $R_{BH}(t) = M_{gal} r_{BH}(t)$
and $S_{\star}(t) = M_{gal} s_{\star}(t)$ so that
\begin{equation}
\label{BHrate}
R_{BH}(t) \simeq 1\times10^{-4} S_{\star}(t)\epsilon_{BH,-3}M_{BH,0,10}^{-1} 
   ~yr^{-1}
\end{equation} 
where $S_{\star}(t)$ is now the star formation rate integrated
over a whole galaxy, expressed in units of solar masses per year.

\section{Accretion onto Stellar Mass Black Holes in Early Galaxies}

After stellar-mass black holes are born, they may grow by accretion
from the early, dense, ISM. This accretion will be subject to 
feedback effects \citep{A09,MBCO09,O10,PR11} and the accretion rate 
will change as the black holes grow in mass. As mentioned in the 
Introduction, the density of the ambient ISM, $n_{ism}$, could have 
been much higher in early galaxies, with some estimates as high as 
$10^{5}cm^{-3}$ \citep{Greif08,MBCO09,PR11,DQMH10,DQM11}. 
The ambient density is, in turn, a principal factor controlling
the accretion.

During the lifetime of a massive main sequence star, stellar winds
will evacuate the local interstellar medium in the vicinity of the 
star. The progenitor star may explode rather than collapsing
quietly to form a black hole. Once the progenitor reaches the end 
of its lifetime and forms a black hole, there will be some time,
$t_{delay}$, as the ISM refills the excavated volume, the bubble
``pops" or the black hole drifts from its natal site. After this delay 
time, the black hole begins accreting from the ambient ISM. The time 
span of the delay may be diminished by neighboring stars exploding as 
supernovae and displacing the ambient gas and dust near the vicinity of
the black hole. Estimates give $t_{delay}\sim10^{8}yr$ 
\citep{Greif08,WA08,Greif10}. The physics governing this delay time 
is complex, but for simplicity we will take the delay time to be 
constant. Over the time spans we will consider, gigayears, this may 
not be an important effect, but we will formally keep the term
through much of the analysis.

Recent computational models have presented accretion rates and
the associated energy production with radiative energy and momentum
feedback for intermediate and supermassive black holes accreting in 
ISM with the high densities expected in young galaxies
\citep{A09,MBCO09,O10,PR11}. The models of \citet{PR11} are
especially useful for our current purpose by providing a 
convenient parameterization of the effects of feedback as a 
ratio with respect to standard Bondi/Hoyle \citep{BH44} accretion. 
\citet{PR11} give the accretion rate for a stationary black hole of 
mass $M_{BH}$ accreting from an isothermal gas of temperature
$T_{ism}$ and density $\rho_{ism}$ as:
\begin{equation}
\label{PR-Bondi}
\dot{M}_{BH,0} = <\lambda_{rad}>_0 \dot{M}_{B,iso,0},
\end{equation}
where the radiative feedback efficiency factor, $<\lambda_{rad}>_0$, is the 
mean ratio of the actual accretion rate to the Bondi/Hoyle accretion 
rate for the same ambient temperature and density, and where the
subscript 0 refers to the assumption that the black hole
has velocity v = 0 with respect to the ambient medium. \citet{PR11}
normalize their results to the Bondi/Hoyle accretion rate
for a stationary black hole in an isothermal gas with sound speed $c_s$
that they take to be:
\begin{equation}
\label{bondi}
\dot{M}_{B,iso,0} = \frac{\lambda_BG^2M_{BH}^2 m_p n_{ism}}{c_{s}^3},
\end{equation} 
where $\lambda_B$ is a coefficient of order unity. Taking $\lambda_B = e^{3/2} \pi = 14.08$
\citep{MBCO09}, Equation \ref{bondi} can be written in convenient units as,
\begin{equation}
\dot{M}_{B,iso,0} = 1.4\times10^{-7}M_{10}^2 n_{ism,5}T_{ism,4}^{-3/2} \rm{\mm yr^{-1}}
\end{equation}
where $M_{10}$ is the black hole mass in units of 10\m, $n_{ism,5}$ 
is the ambient particle density in units of $10^5$ cm$^{-3}$, and $T_{ism,4}$ 
is the ambient temperature in units of $10^4$K, were we have taken 
$T_{ism} = m_p c_{s}^2/k$. For a black hole moving with a velocity, v, 
one would have for the Bondi/Hoyle accretion rate:
\begin{equation}
\label{moving}
\dot{M}_{B,iso} = \dot{M}_{B,iso,0}\frac{c_{s}^3}{(v^2 + c_{s}^2)^{3/2}}.
\end{equation}
The results of \citet{PR11} can be scaled to results appropriate
for a moving black hole by an appropriate scaling of the radiative
feedback efficiency factor and the Bondi/Hoyle accretion rate as
indicated below.

Remarkably, \citet{PR11} \citep[see also][]{MBCO09} find that the 
radiative feedback efficiency parameter, $<\lambda_{rad}>_0$, is independent 
of the the accretion efficiency, $\eta$, essentially independenent of the 
mass of the black hole and insensitive to the ambient density.  They give:
\begin{equation}
\label{rad1}
<\lambda_{rad}>_0 \sim 0.04 T_{ism,4}^{-5/2} \left(\frac{T_{in}}{8\times10^4\rm{K}}\right)^{-4}: 
   n_{ism} > 10^5 \rm{cm^{-3}},  
\end{equation} 
and 
\begin{equation}
\label{rad2}
<\lambda_{rad}>_0 \sim 0.04 T_{ism,4}^{-5/2} \left(\frac{T_{in}}
    {8\times10^4\rm{K}}\right)^{-4}n_{ism,5}^{1/2}: 
   n_{ism} < 10^5 \rm{cm^{-3}},
\end{equation}
where $T_{in}$ is the time-averaged temperature at the accretion radius. 
There is an implicit weak dependence of $<\lambda_{rad}>_0$ on the
mass of the black hole through the parameter, $T_{in}$. The average 
temperature in the inner HII region will be a function of the spectral 
index of the radiation with harder spectral indices giving smaller accretion 
radii and higher $T_{in}$. We are considering smaller mass black holes than did 
\citet{PR11} for which the disk radiation will be harder. To allow for this, we 
take $T_{in}$ corresponding to the hardest spectra considered by \citet{PR11}, 
with spectral index $\alpha = 0.5$ in their Figure 9 and thus $T_{in} \sim 10^5$K. 
We ignore any other scaling of $<\lambda_{rad}>_0$ with $M_{BH}$, although
such dependence may exist. With this scaling, we have:  
\begin{equation}
\label{lambda_rad}
<\lambda_{rad}>_0 \sim 0.016 T_{ism,4}^{-5/2}T_{in,5}^{-4}, 
\end{equation}   
for $n_{ism} > 10^5 \rm{cm^{-3}}$, and a multiplicative factor of $n_{ism,5}^{1/2}$ for 
$n_{ism} < 10^5 \rm{cm^{-3}}$. 

In the final version of their paper, \citet{PR11}
revised Equations \ref{rad1} and \ref{rad2}, reducing the 
overall efficiency from 0.04 to 0.03 and the characteristic 
value of $T_{in}$ to $\sim 4\times10^4$ K. These changes 
would alter the results here by making the overall accretion 
efficiency greater by a factor of $\sim$ 30. Because the value 
of $T_{in}$ for the case we are considering is uncertain, 
we have preserved our normalization to $T_{in} = 10^5$ K. 
Our results can be scaled to other values of $T_{in}$ in the
manner we present, but our result for the luminosity from the
ensemble of stellar mass black holes may be underestimated.
The energy is somewhat less sensitive than the luminosity.
The time, $t_\infty$, for a black hole to grow to infinite
mass might be smaller and more comparable to the initial
accretion delay time, $t_d$, that we have generally neglected.

For the case of a moving black hole, the velocity correction to the accretion
rate as given in Equation \ref{moving} means that for given $T_{ism}$ and
$c_{s}$, the accretion rate given by \citet{PR11} is too large. The
actual accretion rate for a moving black hole would be less, as if the
accretion were occuring at higher effective values of $T_{ism}$ and
$c_{s}$. For a given $T_{ism}$, a moving black hole would accrete
at a rate corresponding to a higher effective ambient temperature of a 
stationary black hole such that: 
\begin{equation}
T_{ism,v=0} = T_{ism,v>0} \frac{v^2 + c_{s}^2}{c_{s}^2},
\end{equation}
where $T_{ism,v>0}$ is the actual ambient temperature and 
$T_{ism,v=0}$ is the ambient temperature (and corresponding sound speed) 
that would give the proper accretion rate in the calculations of \citet{PR11}  
for stationary black holes. If, for a given ambient temperature, the accretion 
rate for a moving black hole corresponds to accretion at effectively a 
higher ambient temperature for a stationary black hole, then the values
of the radiative feedback parameter derived by \citet{PR11} for a given
ambient temperature must also be scaled to that same, higher effective 
temperature or, from Equation \ref{lambda_rad}:
\begin{equation}
<\lambda_{rad}(T_{ism}, v>0)> = <\lambda_{rad}>_0 \left(\frac{v^2 + c_{s}^2}{c_{s}^2}\right)^{-5/2}, 
\end{equation}
where $<\lambda_{rad}>_0$ is the radiative feedback efficiency parameter
for the same value of $T_{ism}$, but for a stationary black hole
from \cite{PR11}. We can thus write for a black hole moving with
a velocity, v:
\begin{equation}
\label{mdot1}
\dot{M}_{BH} = <\lambda_{rad}(T_{ism}, v>0)> \dot{M}_{B,iso} =
   <\lambda_{rad}>_0 \dot{M}_{B,iso,0} \left(\frac{v^2 + c_{s}^2}{c_{s}^2}\right)^{-4}. 
\end{equation} 
The results of \citep{PR11} for a given ambient temperature are thus 
equivalent to those of a somewhat higher ``effective" ambient temperature 
for a moving black hole. This correction factor is quantitatively 
important, but not qualitatively important for $v \lesssim c_{s}$.
The steep dependence on this factor may become significant if the black holes 
move supersonically through the ambient medium. 

With Equations \ref{bondi}, \ref{mdot1}, \ref{rad1} and \ref{rad2}, we can write:
\begin{equation}
\label{mdot2}
\dot{M}_{BH} = F M_{BH}^2, 
\end{equation}
where the factor, F, is:
\begin{equation}
\label{F}
F = <\lambda_{rad}>_0 \left(\frac{v^2 + c_{s}^2}{c_{s}^2}\right)^{-4}
   \frac{e^{3/2}\pi G^2 m_p n_{ism}}{c_{s}^3},   
\end{equation}
or
\begin{equation}
\label{Fnumber}
F = 2.2\times10^{-52} \rm{g^{-1} s^{-1}} T_{ism,4}^{-4} T_{in,5}^{-4} n_{ism,5}^{3/2}
     \left(\frac{v^2 + c_{s}^2}{c_{s}^2}\right)^{-4}; n_{ism,5} < 1,  
\end{equation}
with $c_s^2 = kT_{ism}/m_p$ and the factor of $n_{ism,5}^{3/2}$ would be 
$n_{ism,5}$ for $n_{ism,5} > 1$.  The coefficient, $F$, is a function of the 
ambient medium, but independent of $M_{BH}$ (except for the implicit 
dependence through $T_{in,5}$).
The solution to Equation \ref{mdot2} is thus:
\begin{equation}
\label{MBH1}
M_{BH}(t)=\frac{M_{BH,0}}{\left(1-F M_{BH,0}(t-t_{acc}) \right)},
\end{equation}
where $M_{BH,0}$ is the initial mass of the black hole and 
$t_{acc}$ is the time when the black hole began to accrete. 
For a black hole born at time $t_b$ for which accretion was
delayed by a time $t_{delay}$, then $t_{acc} = t_b + t_{delay}$.
Figure 1 gives a schematic diagram of the growth of a black
hole seed in this framework and Figure 2 gives the black hole
mass as a function of time for various choices of the interstellar
density, with other parameters held constant.

The timescale for a seed black hole to grow to infinite mass is given 
(for $n_{ism,5} < 1$) by:
\begin{equation}
\label{growtime}
t_{\infty} = \frac{1}{F M_{BH,0}} = 
    7.2 Gyr T_{ism,4}^{-4} T_{in,5}^{4} n_{ism,5}^{-3/2}
       \left(\frac{v^2 + c_{s}^2}{c_{s,}^2}\right)^{4}M_{BH,0,10}^{-1}. 
\end{equation}
Note that this time scale increases very steeply with the velocity of the 
black hole.
Writing the velocity correction factor as $f_v = (v^2 + c_{s}^2)/c_{s}^2$ and  
combining Equations \ref{MBH1} and \ref{growtime}, we
can write:
\begin{equation}
\label{MBH2}
M_{BH}(t) = \frac{M_{BH,0}}{\left(1 - \frac{t-t_{acc}}{t_\infty}\right)} =
\frac{M_{BH,0}}{\left(1 - \frac{(t-t_{acc})n_{ism,5}^{3/2}M_{BH,0,10}}{7.2 f_v^4 Gyr}\right)},
\end{equation}
for fiducial temperature parameters, and for $n_{ism,5} < 1$. In our model, the most massive black 
holes will be those that began growing from seeds at a time, $t_{delay}$ after the first 
stellar mass black holes began to form at $t_0$, or $t_{acc} = t_0 + t_{delay}$. With seed
black holes of mass $M_{BH,0} = 10$\m, the most massive black hole will have
a mass of about 14\m\ 2 Gy (at z ~ 3) after the first stellar mass black holes formed 
and about 17\m\ at 3Gy. The mass would formally become infinite at 7.2 Gy for the
chosen parameters, but conditions, especially $n_{ism}$, would have changed by then and 
the accretion might be limited by the Eddington limit. For the timescales of 
interest here and $n_{ism,5} < 1$, this extreme growth is not of 
interest.  We return in \S\ref{Eddington} to discuss the possible effect of growth 
to accretion at the Eddington limit. 

We can now evaluate the number of black holes that will have
formed by a certain epoch, their luminosity at that epoch,
and the energy they will have emitted by that epoch.

The number of black holes of a specific mass, $M_{BH}(t)$ at a time, t, 
will depend on the rate at which the seed black holes were born
and their accretion history. Equation \ref{MBH1} can be inverted to
give the interval of time that a black hole of given mass has
been accreting, $\Delta t_{acc}$:
\begin{equation}
\Delta t_{acc} = t - t_{acc} = \frac{M_{BH}(t) - M_{BH,0}}{F M_{BH,0} M_{BH}(t)}
   = t_\infty\left(1 - \frac{M_{BH,0}}{M_{BH}(t)}\right).
\end{equation}
Since a black hole of given seed mass was born at a time
$t_{b} = t_{acc}-t_{delay} = t - \Delta t_{acc} - t_{delay}$, we can then write 
the time when a black hole of mass $M_{BH}(t)$ was born as:
\begin{equation}
\label{tb}
t_{b} = t - t_\infty\left(1 - \frac{M_{BH,0}}{M_{BH}(t)}\right) - t_{delay},
\end{equation} 
from which we can write:
\begin{equation}
\label{dtb}
\frac{dt_{b}}{dM_{BH}} = -\frac{1}{F M_{BH}^{2}(t)}.
\end{equation}
Note the sign change here with respect to Equation \ref{mdot2} that results
from the difference in taking the derivative with respect to t, holding
$t_b$ constant versus taking the derivative with respect to $t_b$ holding
the currrent epoch, t, constant.
The rate at which black holes of mass $M_{BH}(t)$ at time t were born is
the rate at which their seeds were born at time, $t_b$, which can thus
be expressed as:
\begin{equation}
R_{BH}(t_b) = R_{BH}(t - t_\infty\left(1 - \frac{M_{BH,0}}{M_{BH}(t)}\right) - t_{delay}).
\end{equation}
Since black holes of larger current mass were born earlier, black holes
with mass between $M_{BH}$ and $M_{BH}$ + d$M_{BH}$ were born between
$t_{b}$ and $t_{b}$ - d$t_{b}$. At current epoch, t, the number of black holes 
with mass between $M_{BH}$ and $M_{BH}$ + d$M_{BH}$ is thus given by: 
\begin{equation}
\label{diffnumber}
\frac{dN_{BH}(t)}{dM_{BH}}dM_{BH} = \frac{dN_{BH}(\Delta t_{acc})}{d\Delta t_{acc}}
   \frac{d\Delta t_{acc}}{dM_{BH}}dM_{BH}
      = \frac{dN_{BH}(t_b)}{dt_b}\frac{\Delta t_{acc}}{dt_b}\frac{dt_b}{dM_{BH}}dM_{BH},
\end{equation}
where we have taken $\frac{dN_{BH}(\Delta t_{acc})}{d\Delta t_{acc}} = \frac{dN_{BH}(t_b)}{dt_b}$.
With $\frac{dN_{BH}(t_b)}{dt_b} = R_{BH}(t_b)$, $\frac{\Delta t_{acc}}{dt_b} = -1$,
and Equation \ref{dtb}, we can then write:   
\begin{equation}
\frac{dN_{BH}(t)}{dM_{BH}}dM_{BH} = R_{BH}(t_b)\frac{dM_{BH}}{F M_{BH}(t)^2}.
\end{equation}

The number of black holes in a galaxy with a given total star formation 
rate and hence total black hole formation rate born between the beginning 
of the epoch when stellar-mass black holes formed, $t_0$, and the epoch under 
consideration, $t$, is thus: 
\begin{equation}
\label{NBH}
N_{BH}(t) = \int_{t_0}^{t} R_{BH}(t')dt' = 
   \int_{M_{BH,0}}^{M_{BH,max}(t)}\frac{R_{BH}(t_{b})}{F M_{BH}(t')^2}dM_{BH} 
      + \int_{t -t_{delay}}^{t} R_{BH}(t')dt',
\end{equation}
where $t_b(t')$ is given by Equation \ref{tb} and   
\begin{equation}
\label{mbhmax}
M_{BH,max}(t) =  \frac{M_{BH,0}}{1-(t-t_0-t_{delay})/t_{\infty}}, 
\end{equation}
is the maximum mass to which a black hole could have grown at time t,
one which began accreting as soon as it could, a time $t_{delay}$
after the first stellar-mass black holes were born at $t_0$. The 
second integral over $R_{BH}$ on the right hand side 
accounts for the black holes
that were born during the last interval, $t_{delay}$, before the epoch, t,
under consideration that have not yet begun to accrete. The integral of 
$R_{BH}$ over the full time interval, $t_0$ to $t$, in Eqn \ref{NBH} 
yields the desired result for the number of black holes (especially in 
the trivial case for which $R_{BH} = constant$), but the first integral 
on the right hand side provides the framework for evaluating the luminosity 
and energy production, as described below. There is no additive term over
the last interval of $t_{delay}$ in the computation of luminosity and energy, 
since these late-born black holes are, by assumption, not yet accreting.

In principle, to evaluate the integral in Equation \ref{NBH} one needs 
to take into account the inhomogeneity and temporal variations in the 
ambient quantities that determine the factor, $F$, especially $n_{ism}$
and the velocity term, $f_v$, and the temporal variation in the formation 
rate of black holes that itself is determined by variations in the star 
formation rate (Equation \ref{BHrate}). When young galaxies frequently 
collide, the star formation rate is expected to be very bursty \citep{DQM11}. 
To get a qualitative feel for the expectations of our framework, we will 
take the ambient conditions and the star formation rate to be constant in 
time. A representative star formation rate for a fiducial galaxy of mass 
$M_{gal} = 4\times10^{12}$ \m\ from the merging galaxy simulations of 
\citet{DQM11} is $S_{\star} \sim 10$ \m~$yr^{-1}$ and hence from 
Equation \ref{BHrate}, $R_{BH} \sim 1\times10^{-3}\epsilon_{BH,-3}M_{BH,0,10}^{-1}
\left(\frac{S_{\star}}{10\mm yr^{-1}}\right)$ $yr^{-1}$.
We can then write:
\begin{eqnarray}
N_{BH}(t) &\sim& -\frac{R_{BH}}{F} 
    \int_{M_{BH,0}}^{M_{BH,max}(t)}\frac{dM_{BH}}{M_{BH}(t)^2} \nonumber \\
         &=& R_{BH}\left(t - t_0 \right) 
         \nonumber \\ 
         &\sim& 1\times10^6 t_{Gyr} \epsilon_{BH,-3}
                    M_{BH,0,10}^{-1}\left(\frac{S_{\star}}{10\mm yr^{-1}}\right),
\end{eqnarray}  
where in the last step we have taken t to be measured in Gyr and have neglected
$t_0$ compared to timescales of Gyr. We have taken the more cumbersome means to 
compute this quantity from Eqn. \ref{NBH} for illustration, but note that the 
result is consistent with the simple integration over the rate of production of 
black holes. We thus estimate that over a period of 2 Gy, a galaxy like the Milky
Way could have produced of order $2\times10^6$ black holes that had grown from 
seeds of 10 \m\ to various masses. Figure 3 gives the number of black holes as 
a function of time for various choices of the initial black hole mass, $M_{BH,0}$,
with other parameters held constant.

We can now use the same framework to estimate the luminosity of the 
ensemble of accreting stellar-mass black holes at the epoch, t.   
We will adopt the parameterization of \citet{PR11} with the assumption 
that there is sufficient angular momentum to form a disk near the 
black hole, so that disk-like efficiencies for turning mass accretion 
rates into radiated energy are applicable. For critiques of
this assumption, see \citet{RA94,BK05}. We write for the luminosity
of a single accreting black hole:
\begin{equation}
\label{Lsingle}
L(t) = \eta \dot{M}_{BH}(t)c^2,
\end{equation} 
with $\eta \sim 0.1$. If a disk does not form, or
forms only sporadically, the radiation efficiency would be 
correspondingly less than the fiducial value we assume here. 

The luminosity per unit galaxy mass radiated by all black holes
with mass between $M_{BH}$ and $M_{BH} + dM_{BH}$ at epoch, t,
is given by: 
\begin{equation}
\frac{dL_{tot}(t)}{dM_{BH}(t)}dM_{BH}
   =\frac{dN_{BH}(t)}{dM_{BH}(t)}dM_{BH}\eta c^{2}F M_{BH}^{2}(t).
\end{equation}
Using Equation \ref{diffnumber}, this can be written as:
\begin{equation}
\frac{dL_{tot}(t)}{dM_{BH}(t)}dM_{BH} = R_{BH}(t_b)\eta c^{2}dM_{BH}.
\end{equation}
The total luminosity from all the accreting black holes born
since $t_o$ can then be obtained by integrating over all the current
masses at epoch, t, to obtain:
\begin{equation}
L_{tot}(t) = \int_{M_{BH,0}}^{M_{BH,max}(t)}R_{BH}(t_{b})\eta c^{2}dM_{BH}.
\end{equation}

Once again, we can approximate the complex variation of the rate
of production of black holes with a constant to obtain:
\begin{equation}
\label{luminosity}
L_{tot}(t) \sim R_{BH}\eta c^{2}(M_{BH,max}(t)-M_{BH,0})
   = R_{BH}\eta c^{2} M_{BH,0} \left(\frac{t - t_0 - t_{delay}}
        {t_{\infty} + t_0 + t_{delay} -t}\right).
\end{equation}
Neglecting $t_0$ and $t_{delay}$, taking $t << t_\infty$, and
using $N_{BH} \sim R_{BH}t$ gives a simple estimate of the 
luminosity of $L_{tot} \sim N_{BH}(t) \eta c^2 M_{BH,0}/t_\infty$. 
Neglecting $t_0$ and $t_{delay}$ we have for the total luminosity of a 
galaxy of constant star formation rate:
\begin{equation}
L_{tot}(t) \sim \frac{8\times10^{42} {\rm erg s^{-1}} t_{Gyr}}
      {1 - 0.14 t_{Gyr}n_{ism,5}^{3/2}f_v^{-4}M_{BH,0,10}}
        \epsilon_{BH,-3}n_{ism,5}^{3/2}f_v^{-4}M_{BH,0,10}
           \left(\frac{S_*}{10\mm yr^{-1}}\right), 
\end{equation}
where we have taken the timescale $t_{\infty}$ 
from Equation \ref{growtime} to be $7.2 n_{ism,5}^{-3/2}f_v^4M_{BH,0,10}$ 
Gyr for fiducial temperature parameters and for $n_{ism,5} < 1$. 
For $t \sim 2$ Gyr, the ensemble of about $10^6$ black holes
will produce $\sim 10^{43}$ erg s$^{-1}$, as much luminosity as 
a modest Seyfert galaxy or a single black hole of about $10^5$\m\
accreting near the Eddington limit. For comparison, supernovae provide 
an input of about $L_{SN} \sim 10^{49}$ erg yr$^{-1}$ in a galaxy like
the Milky Way for which the star formation rate is about 1 \m\ yr$^{-1}$,
so a star formation rate of 10 \m\ yr$^{-1}$ might give $L_{SN} \sim
3\times10^{42}$ erg s$^{-1}$. The black hole input at about 2Gy is
comparable to, and might even slightly exceed the input power from
supernovae. Figure 4 gives the luminosity versus time for various 
values of the interstellar density and Figure 5 illustrates the 
sensitivity of the luminosity to the velocity parameter, $f_v$.

The total energy liberated by the number of black holes, 
$N_{BH}$, accreting from $t_o + t_{delay}$ to $t$ is 
\begin{equation} 
E_{tot} = \int_{t_0 + t_{delay}}^{t}L_{tot}(t')dt' = 
    \int_{t_0 + t_{delay}}^{t} \int_{M_{BH,0}}^{M_{BH,max}(t')}R_{BH}(t_{b})\eta c^{2}dM_{BH}.
\end{equation}
We will again adopt the approximation of a constant value of $R_{BH}$ to write
\begin{equation}
E_{tot} \sim R_{BH}\eta c^2 \int_{t_0 + t_{delay}}^{t} (M_{BH,max}(t') - M_{BH,0})dt'.
\end{equation}
Invoking
\begin{equation}
\frac{dM_{BH,max}(t')}{dt'} = F M_{BH,max}^2(t'),
\end{equation}
we can write,
\begin{equation}
E_{tot}(t) = \frac{R_{BH}\eta c^{2}}{F} 
    \int_{M_{BH,0}}^{\frac{M_{BH,0}}{1 - FM_{BH,0}(t - t_0 - t_{delay})}}
        \frac{M_{BH,max} - M_{BH,0}}{M_{BH,max}^2}dM_{BH,max},
\end{equation}
which becomes 
\begin{equation}
E_{tot}(t) = R_{BH} \eta c^{2} M_{BH,0} t_\infty
     \left\{-ln\left(1-\frac{t - t_0 - t_{delay}}{t_\infty}\right) + 
           \frac{t - t_0 - t_{delay}}{t_\infty}\right\}. 
\end{equation}
For $t_0 + t_{delay} << t << t_{\infty}$, this reduces to 
\begin{equation}
E_{tot} \sim 2R_{BH}\eta c^2 M_{BH,0} 
   \left(t + \frac{1}{4}\frac{t^2}{t_\infty}\right) 
     \sim 2L_{tot}(t)t_{\infty} \sim 2N_{BH}(t)\eta c^2 M_{BH,0}, 
\end{equation}
nearly independent of $t_{\infty}$. 
Again taking $t_{\infty}$ from Equation \ref{growtime} to be 
$7.2 n_{ism,5}^{-3/2}f_v^4M_{BH,0,10}^{-1}$ Gyr for fiducial temperatures
and neglecting $t_0$ and $t_{delay}$ on the timescales of interest, we have:
\begin{eqnarray}
E_{tot}(t)&\sim&4.1\times10^{71} \rm{ergs}~R_{BH} n_{ism,5}^{-3/2}f_v^4
    \nonumber \\
    &&\times\left\{-ln\left(1- 0.14t_{Gyr}n_{ism,5}^{3/2}f_v^{-4}M_{BH,0,10}\right) + 
        0.14 t_{Gyr}n_{ism,5}^{3/2}f_v^{-4}M_{BH,0,10}\right\}.
\end{eqnarray}
For $R_{BH} \sim 3.2\times10^{-11} \rm{s}^{-1}\epsilon_{BH,-3}M_{BH,0,10}^{-1}
        \left(\frac{S_*}{10\mm yr^{-1}}\right)$, we have
\begin{eqnarray}
E_{tot}(t)&\sim&1.3\times10^{60} \rm{ergs}~\epsilon_{BH,-3}M_{BH,0,10}^{-1}
           \left(\frac{S_*}{10\mm yr^{-1}}\right)
            n_{ism,5}^{-3/2}f_v^4 \nonumber \\
     &&\times\ \left\{-ln\left(1- 0.14t_{Gyr}n_{ism,5}^{3/2}f_v^{-4}M_{BH,0,10}\right) + 
        0.14 t_{Gyr}n_{ism,5}^{3/2}f_v^{-4}M_{BH,0,10}\right\}.
\end{eqnarray}
For $t \sim 2$ Gyr after the beginning of stellar-mass black hole formation, 
corresponding to a redshift of about 3, we have
\begin{eqnarray}
E_{tot}(t) &\sim& 1.3\times10^{61} \rm{ergs}~\epsilon_{BH,-3}M_{BH,0,10}
           \left(\frac{S_*}{10\mm yr^{-1}}\right)
            n_{ism,5}^{-3/2}f_v^4 \nonumber \\
     &&\times\ \left\{-ln\left(1- 0.28n_{ism,5}^{3/2}f_v^{-4}M_{BH,0,10}\right) + 
        0.28 n_{ism,5}^{3/2}f_v^{-4}M_{BH,0,10}\right\}.
\end{eqnarray}
For a density of $n_{ism} = 10^5$ cm$^{-3}$, an initial black hole mass of
10 \m\ and taking the velocity factor, $f_v$ to be unity, we get an energy liberated
of $E_{tot}(t) \sim 8\times10^{60} \rm{ergs}$ by 2 Gyr. This is equivalent to
$8\times10^{9}$ supernovae liberating about $10^{51}$ ergs apiece, but
the energy would be entirely in radiant, not kinetic energy, in our model. In
practice, if the black holes are accreting by means of an accretion disk, as
we assume, then some of the energy emitted may be in the form of jets
and hence of kinetic energy. Figure 6 gives the total liberated energy versus 
time for various values of the interstellar density and other parameters held 
constant. Figure 7 gives the total energy for various choices of the
velocity parameter, $f_v$ at a density of $n_{ism,5} =1$. Figure 8 gives
the total energy as a function of the parameter, $t_\infty$.  

\section{Eddington Limit Concerns}
\label{Eddington}

In the previous discussion, we have assumed accretion given by
the Bondi/Hoyle rate with radiation feedback. Each seed black hole
will accrete more rapidly with time, and there can be an epoch
when the seed black holes reach the Eddington limit (for the
relevant opacity). In this case, the accretion would proceed 
in a different manner. 

We can solve for the time at which the Eddington limit for 
electron scattering is reached during the Bondi accretion phase 
by setting the accretion luminosity for a single black hole from 
Equation \ref{Lsingle} equal to the Eddington luminosity: 
\begin{equation}
L(M_{BH}(t)) = \eta FM_{BH}^{2}c^{2} = L_{Edd} = KM_{BH}
\end{equation} 
where $K{\approx}10^{5}$ erg s$^{-1}$ g$^{-1}$.
We thus have for the black hole mass at the Eddington limit, $M_{BH,Edd}$, 
\begin{equation}
M_{BH,Edd} = \frac{K}{\eta Fc^{2}} \sim 250~\mm\ n_{ism,5}^{-3/2}f_v^4, 
\end{equation}
for fiducial values of temperature parameters in Equation \ref{Fnumber}. 
The time when the first seed reaches the Eddington limit is obtained by equating
$M_{BH,Edd}$ with $M_{BH,max}$ from Equation \ref{mbhmax}, neglecting
the delay time, or,
\begin{equation}
M_{BH,Edd} = \frac{K}{\eta Fc^{2}} = M_{BH,max} = 
    \frac{M_{BH,0}}{1-F M_{BH,0}(t_{Edd}-t_0)}, 
\end{equation}
or,
\begin{equation}
t_{Edd} -t_0 = \frac{K - \eta Fc^{2}M_{BH,0}}{KFM_{BH,0}} 
     \sim t_\infty, 
\end{equation}
where in the final step we have neglected the second
term in the numerator that is numerically small compared to $K$.
Thus, formally we would have for $M_{BH} < \frac{K}{{\eta}Fc^2}$: 
\begin{eqnarray}
\dot{M}_{BH} &=& FM_{BH}^2 \nonumber \\
     M_{BH} &=& \frac{M_{BH,0}}{1-F(t-t_{acc})M_{BH,0}} \nonumber \\
         L(t) &=& {\eta}Fc^2M^2_{BH},
\end{eqnarray}
and for $M_{BH} > \frac{K}{{\eta}Fc^2}$:
\begin{eqnarray}
\dot{M}_{BH} &=& \frac{KM_{BH}}{{\eta}c^2} \nonumber \\
    M_{BH} &=& \frac{K}{{\eta}Fc^2}e^{\frac{Kt}{\eta c^2}} \nonumber \\
      L &=& KM_{BH}.
\end{eqnarray}
In practice, $t_{Edd} \sim t_\infty$ is so long even under conditions 
of high density, $n_{ism} \sim 10^5$ cm$^{-3}$, that it is 
unlikely that any of the seed black holes would reach the 
Eddington limit before the ambient conditions had changed to  
render the Eddington limit even further out of reach. For
the conditions we envisage, the Eddington limit has no
practical influence on the growth and radiation from 
the ensemble of stellar mass black holes.

\section{Discussion and Conclusions}

A plethora of stellar-mass black holes accreting in young galaxies
could be significant source of radiant energy. Unlike the feedback
from stars and supernovae that are one-time, rather short term
events in the history of a galaxy, individual stellar-mass 
black holes accumulate and continue to radiate as long as the 
conditions promote accretion. A principal factor is the ambient 
density of the ISM that must remain sufficiently high.  We have shown
that for predicted ambient densities in the range $10^4 - 10^5$ cm$^{-3}$,
the condition for appreciable accretion is met in the context of
Bondi/Hoyle accretion limited by radiative feedback. In this case, 
there is potentially a significant amount of accretion-radiated energy 
available. 

We derive the total energy production of stellar mass black holes
accreting from a high-density primordial ISM using the classical Bondi
spherically symmetric accretion rate scaled with the value for the
mean accretion rate affected by feedback given in \citet{PR11}. Since
stellar mass black holes will be distributed in proportion to star formation,
their energy production will be emitted more uniformly over the galaxy
than a single supermassive black hole. This may have implications
for early galaxy evolution as large scale galaxies start to form during
the era of reionization. We also show that given the estimated values
used in our calculations, the stellar-mass black holes are not likely to
reach the Eddington limit during the time span we have used here,
a few Gyr.

For our analytic calculations, the ISM density, $n_{ism}$, has been 
taken to be constant in space and over the time spans of interest. 
In reality, the density will vary througout the galaxy and as the galaxy 
evolves over time, the density will decrease. 
The motion of the black holes with respect to the background
gas could also be an important limiting factor as the motion approaches
or exceeds supersonic. This motion could affect the nature of the
feedback process itself, but, to the best of our knowledge, this
has not been explored in any detail.  Due to these effects, our 
calculations may be more optimistic than a more realistic case that
accounts for density fluctuations and evolution and for black
hole motion. Some account of 
these factors might be taken by replacing the density of the ISM
in our formulation, $n_{ism}$ with $f_f n_{ism}$ where $f_f$ is a filling
factor corresponding to a given density. The ambient density may be
affected by whether star formation is dominated by mergers or 
by cold inflow at the epochs of interest \citep{S08}. We have, however, 
given a general framework by which the accumulating number of 
stellar-mass black holes, their luminosity at a given epoch, and
their accumulated feedback energy could be incorporated in a 
numerical simulation that followed the fluctuations and density
evolution more realistically. A simulation that included the
distributed feedback effects of accumulating, accreting
stellar-mass black holes might give significantly different results
than one that only included stars, supernovae and one or a few
intermediate-mass or supermassive black holes. 

As we were finishing work on this paper, the paper by 
\citet{Mirabel11} appeared that considers the possible role of
stellar-mass black holes in high-mass X-ray binaries in 
young galaxies. This scenario has the advantage that the
accretion is driven by mass transfer from a companion star
and does not depend on the vagaries of the ISM. On the other
hand, this proposition does depend on the binary fraction
of massive stars in the early Universe, a rather uncertain
quantity. \citet{Mirabel11} seem to assume that all stellar-mass
black holes formed in these early epochs are in binary systems.
We note that these high-mass X-ray binaries live for a short
time \citep[perhaps 0.02 Gyr,][]{Mirabel11} and hence each binary
system is a delta-function contribution on the time scales we 
consider here. When the secondary dies, the primary
black hole (or a secondary one if it forms) would then
be subject to the sort of radiative-feedback limited
Bondi accretion we consider here. Both single and
binary black holes should be considered in complete
models of the evolution of young galaxies.  

While we have concentrated on epochs $\sim$ Gyr, the existence
and impact of stellar-mass black hole accretion and feedback
may be pertinent to the earliest phases of star and galaxy 
formation if, for instance, turbulence allows the formation
of a broader initial mass function and hence lower mass 
stars already in the first dark matter halos where star 
formation is thought to have first begun \citep{Clark11,Prieto11}. 
Whereas a single very massive star explosion by pair-instability 
releasing $\sim 10^{53}$ ergs is likely to unbind a mini-halo, the
explosion of less massive ``normal" supernovae and the 
accretion and feedback of stellar-mass black holes may yield
a different behavior than many current simulations envisage.
Supernovae may blow fountains rather than totally disrupting
the baryonic content of the halos and feedback from accreting
black holes may alter state of ionization of the gas. Both
types of events may drive further turbulence. If stellar-mass
black holes are present from the first epochs of star formation,
then their role must be considered in the subsequent mergers
that build more massive galaxies. Aside from the effects of 
accretion and feedback, any stellar-mass black holes that 
form and remain in mini halos may themselves merge through 
dynamical friction and help to promote the growth of larger-mass
black holes during the era of rapid galaxy merging. The
tendency to move supersonically with respect to the ambient
gas that will tend to limit the accretion may promote this 
dynamical friction and merging.   

An important aspect that we have not yet explored in depth is 
the observational consequence of our model. This might be
addressed by a simulation incorporating stellar-mass black 
hole feedback. A related issue might be to discriminate 
the signal of a host of stellar-mass black holes from single,
larger-mass black holes. One means to do this might be to
make both radio and X-ray observations of young galaxies. 
There appears to be a nearly universal relation between
the radio and X-ray luminosity of black holes such that
$L_{radio} \propto L_X^{0.7}$, where both luminosities are
presumed to scale monotonically with black hole mass. If
this holds for the conditions we address here, then an
ensemble of stellar-mass black holes should have a
larger radio flux for a given X-ray flux than a single
black hole of the same total mass. In this context, 
\citet{Jia11} have examined the evidence for the growth
of black holes in local starburst galaxies. They find
an anomalous increase in X-ray flux that they argue could
be associated with the growth of a black hole of $10^4$\m\
by accretion. We suggest there might be an ensemble of 
$10^3$ smaller-mass black holes that might provide the
same X-ray flus, in which case the radio luminosity would
be relatively large.  

Combining Eqns \ref{mdot2} and \ref{Fnumber} we estimate
the accretion rate for fiducial parameters to be $\sim
10^{-9}$ \m\ yr$^{-1}$. This accretion rate corresponds
roughly to that needed to induce the accretion disk
limit-cycle instability \citep{CGW82,LFP85} and thus to
produce a black hole X-ray nova analogous to AO620-00
and related events \citep{TL95}. The X-ray flux from these
systems might thus come in flares near the Eddington
limit lasting for months with quiescent periods of
years to decades. In an active star-forming galaxy at a 
red shift of about 2 with $\sim 10^6$ stellar mass black 
holes, $\sim10^4$ of them might be in outburst at any given 
time. Any such bursts may be redshifted into bands that are 
heavily extincted and hence difficult to observe directly, but
this aspect is worth considering more carefully. 

\citet{K11} have argued that black holes do not correlate 
with disks and that they correlate little, if at all, with 
pseudobulges. It would be interesting to consider whether
or not stellar-mass black holes affect the early evolution
of disk-grown pseudobulges.

There is currently a problem understanding the rapid
early growth of dust in young galaxies. Perhaps the 
ensemble of stellar-mass black holes contributes to dust 
formation by, for instance, providing numerous and prolonged 
concentrations of ISM density at the boundary of radiative 
feedback regions.

Finally we note that our estimates here suggest that a contemporary
stellar mass black hole that found itself in a dense clump of 
molecular material, $n \sim 10^5$ cm$^{-3}$, might be observably luminous
$\sim 10^{37}$ erg s$^{-1}$ for fiducial parameters at this density.
Dense clumps like this are estimated to fill a volume in the Galaxy
of about $4\times10^4$ pc$^3$ (N. J. Evans, private communication). With 
a volume within 100 pc of the Galactic plane of $\sim 10^{11}$ pc$^3$, 
this represents a filling factor for dense molecular clumps of 
$\sim 4\times10^{-7}$. With estimates of $10^6 - 10^8$ black holes in 
the Galaxy, the estimated number of black holes within such dense 
clumps ranges from negligible to a few.

\acknowledgments
We thank Milos Milosavljevic, Pawan Kumar, Volker Bromm,
Jenny Greene, and Neal Evans for useful discussions. Special 
thanks go to Tracey Bennett who helped with the literature search.
VJ is grateful for an award from the College of Natural
Sciences and for support from the Undergraduate Excellence
fund of the Department of Astronomy. This work was supported 
in part by NSF Grant AST-0707769.

{}


\newpage

\begin{figure}[htp]
\centering
\includegraphics[totalheight=0.5\textheight]{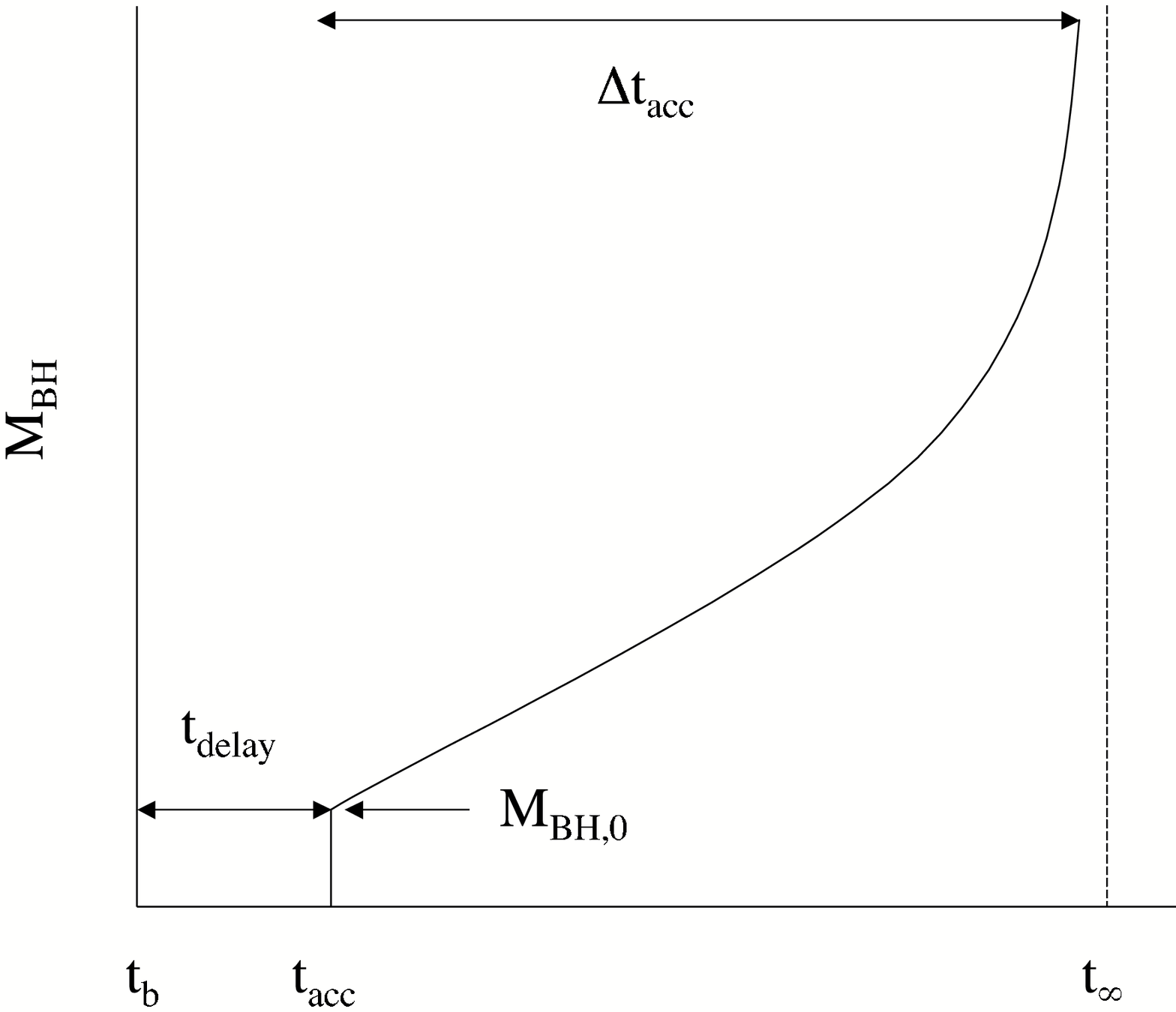}
\figcaption[Mschematic.ps]{Schematic diagram of the growth
of a black hole from a seed by feedback-limited 
Bondi/Hoyle accretion in an early galaxy.}
\end{figure}

\newpage

\begin{figure}[htp]
\centering
\includegraphics[totalheight=0.5\textheight]{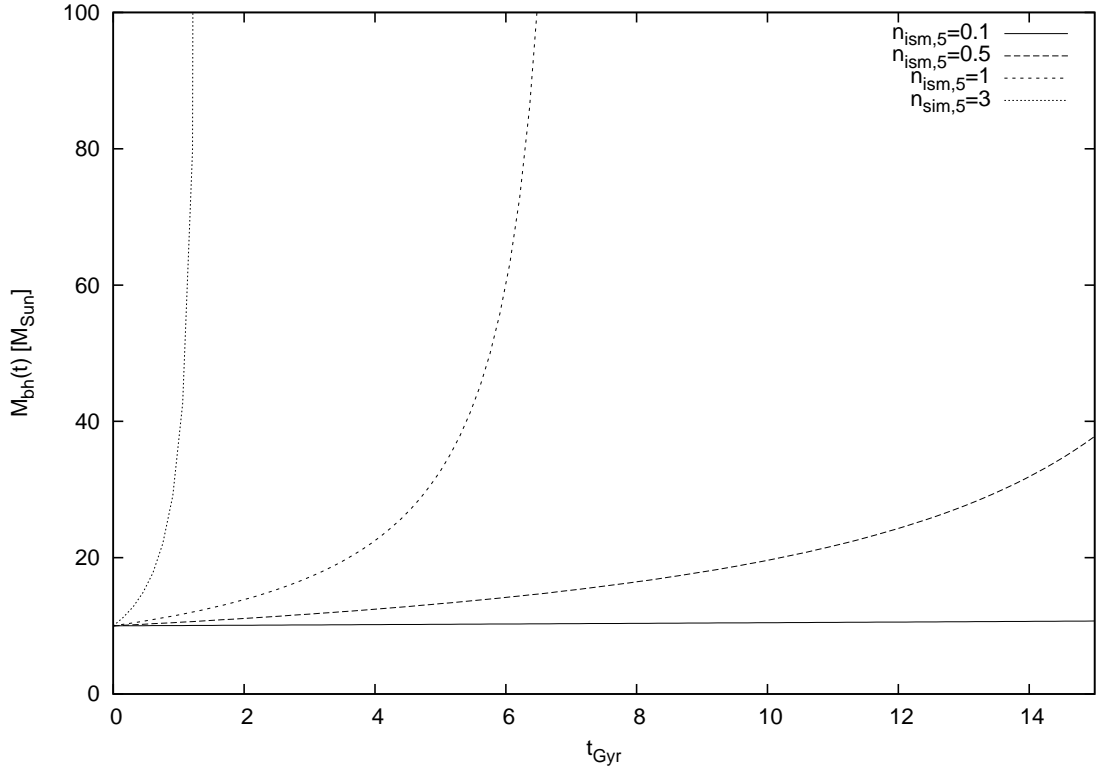}
\figcaption[Mbh.ps]{The mass of a stellar-mass black hole as a 
function of the time the black hole accretes for an initial mass 
of 10 \m\ in an early galaxy with star formation rate 
10 \m\ yr$^{-1}$ for various values of the ambient density, $n_{ism}$.
Other parameters are set to fiducial values (see text).}
\end{figure}

\newpage

\begin{figure}[htp]
\centering
\includegraphics[totalheight=0.5\textheight]{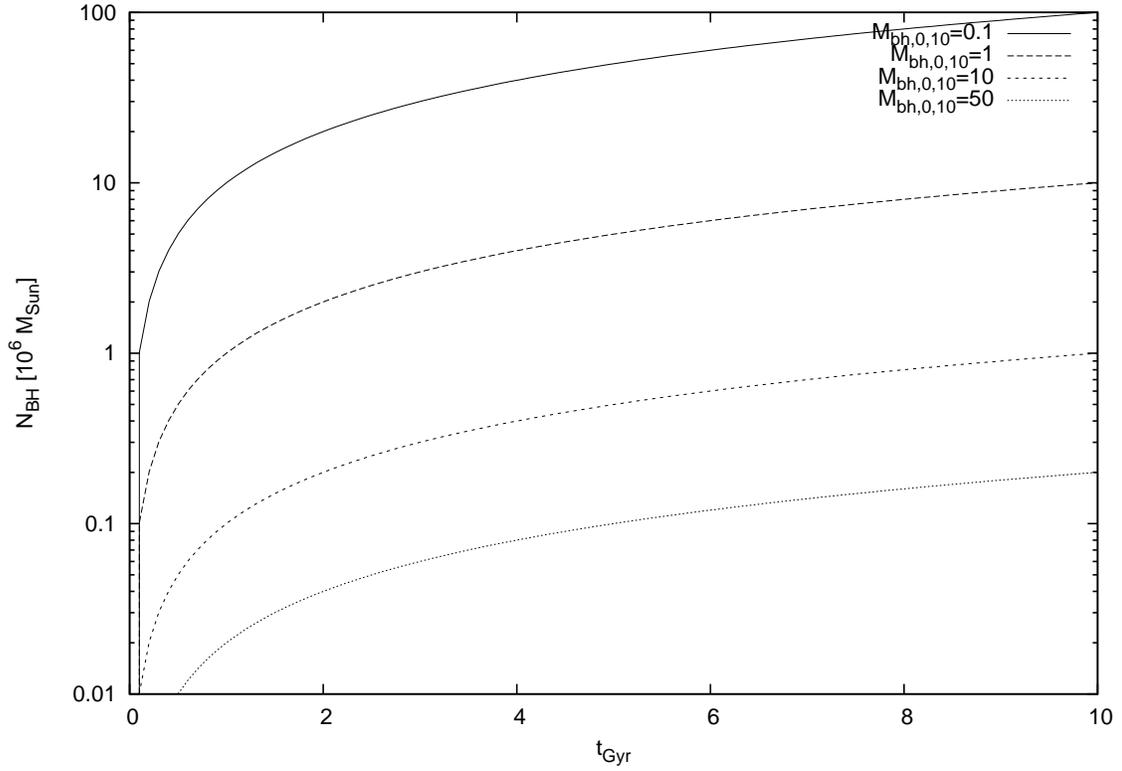}
\figcaption[Nbh.ps]{The number of stellar-mass black holes as a 
function of time in an early galaxy with star formation rate 
10 \m\ yr$^{-1}$ for various values of the initial black hole mass, $M_{BH,0}$.
Other parameters are set to fiducial values (see text).}
\end{figure}

\newpage

\begin{figure}[htp]
\centering
\includegraphics[totalheight=0.5\textheight]{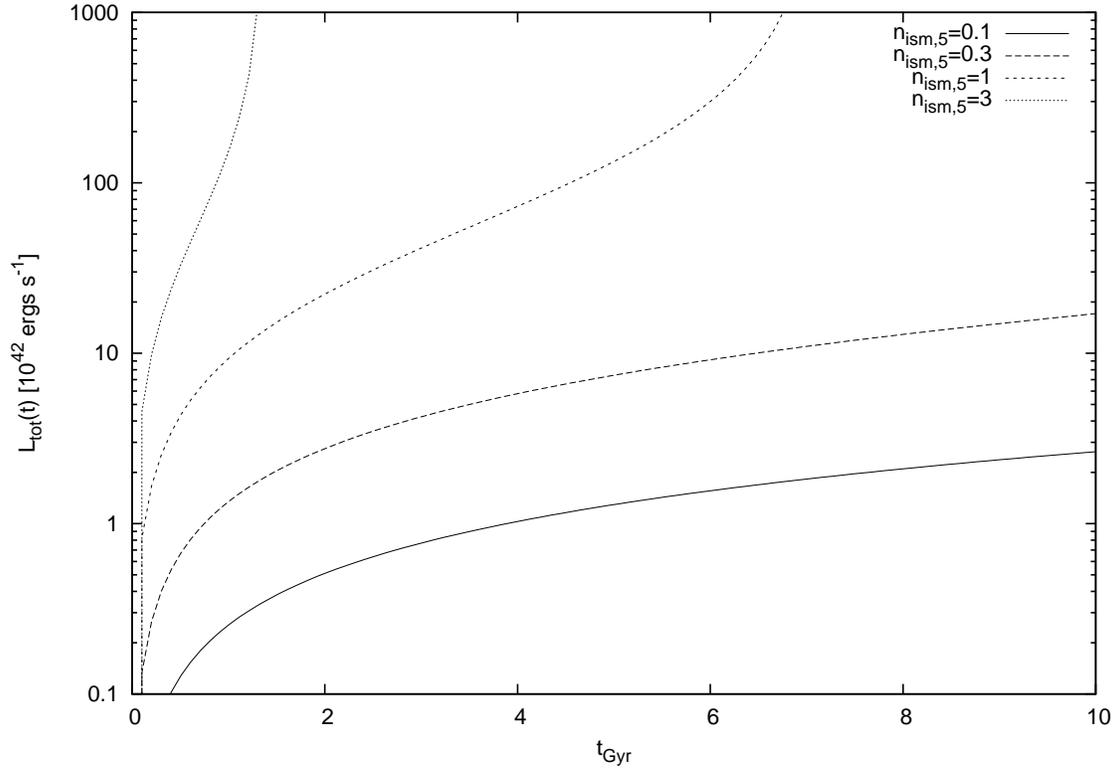}
\figcaption[Lbh_n.ps]{The luminosity of stellar-mass black holes
as a function of time in an early galaxy with star formation rate
10 \m\ yr$^{-1}$ for various values of the ambient density, $n_{ism}$.
Other parameters are set to fiducial values (see text).}
\end{figure}

\newpage

\begin{figure}[htp]
\centering
\includegraphics[totalheight=0.5\textheight]{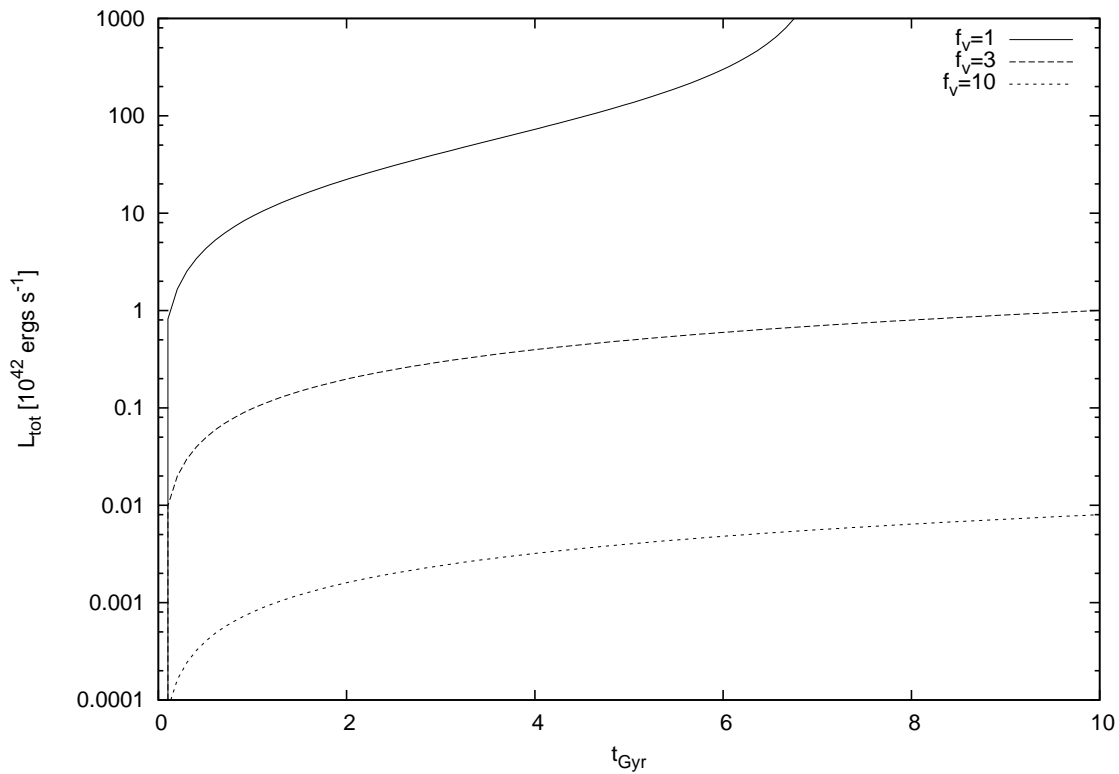}
\figcaption[Lbh_f.ps]{The luminosity of stellar-mass black holes
as a function of time in an early galaxy with star formation rate
10 \m\ yr$^{-1}$ for various values of the velocity parameter
$f_v = (v^2 + c_s^2)/c_s^2$ and $n_{ism} = 10^5$ cm$^{-3}$.
Other parameters are set to fiducial values (see text).}

\end{figure}

\newpage

\begin{figure}[htp]
\centering
\includegraphics[totalheight=0.5\textheight]{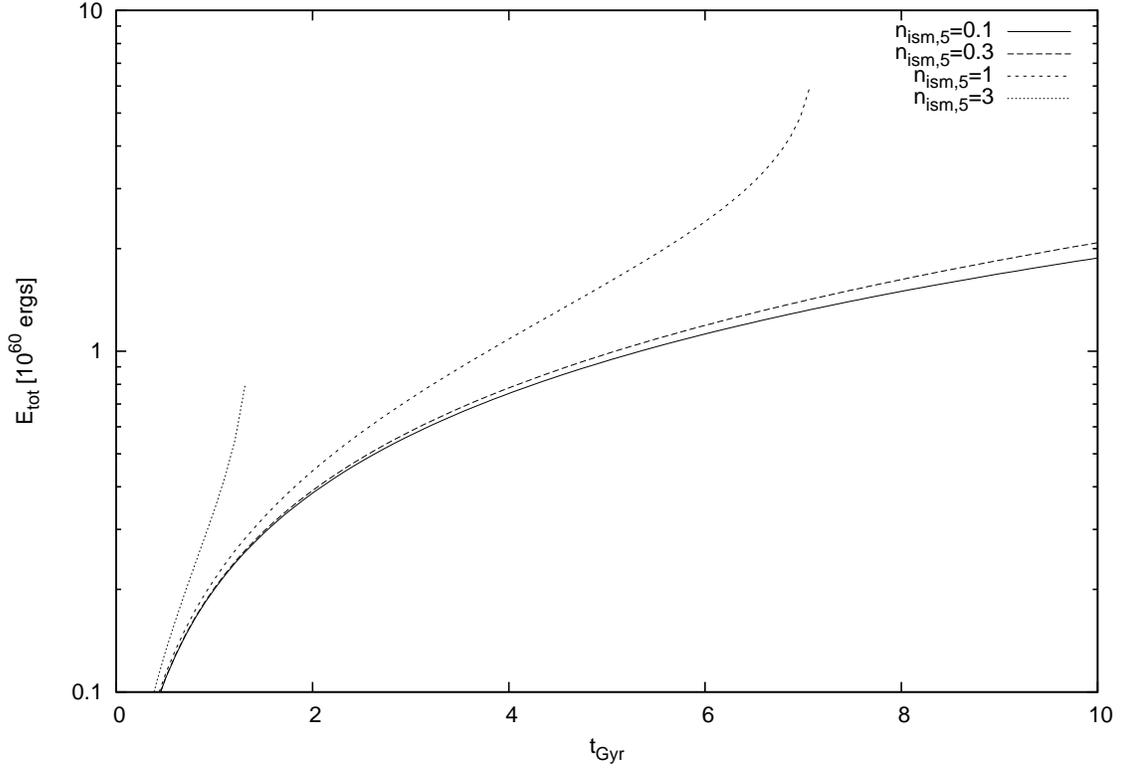}
\figcaption[Ebh_n.ps]{The total integrated energy of stellar-mass 
black holes as a function of time in an early galaxy with star formation 
rate 10 \m\ yr$^{-1}$ for various values of the ambient density, $n_{ism}$.
Other parameters are set to fiducial values (see text).
The curves are truncated for t approaching t$_\infty$ from below 
due to graphics resolution.
}
\end{figure}

\newpage

\begin{figure}[htp]
\centering
\includegraphics[totalheight=0.5\textheight]{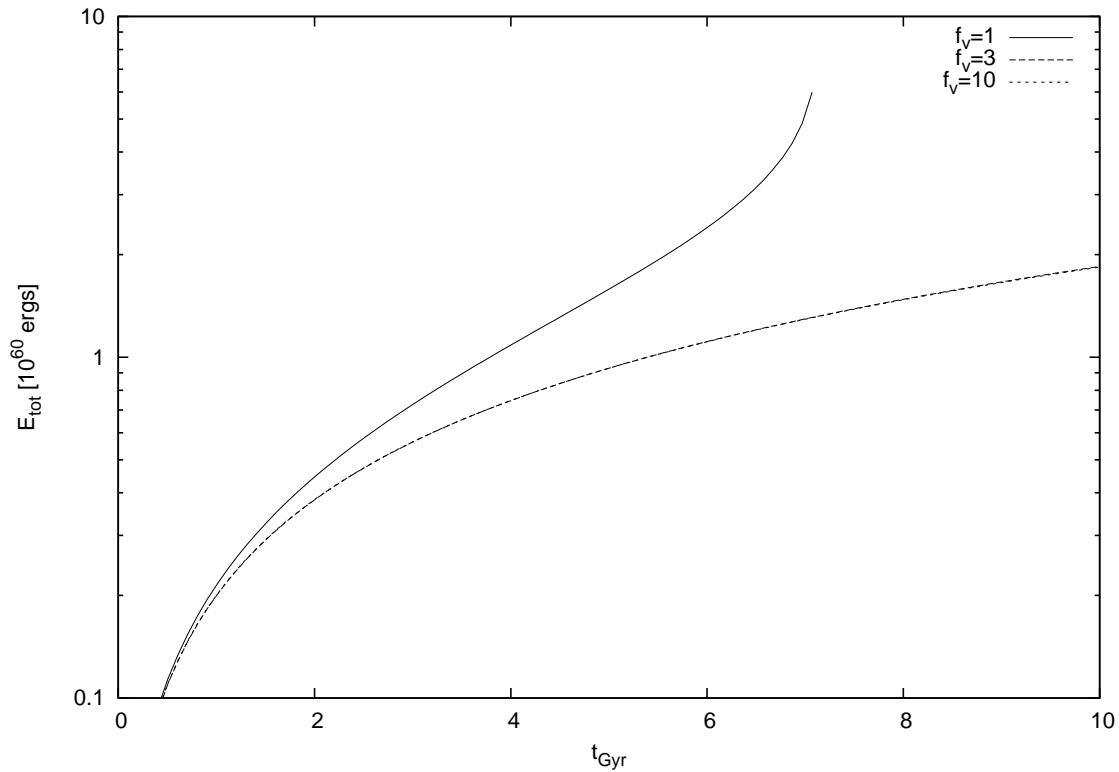}
\figcaption[Ebh_f.ps]{The total integrated energy of stellar-mass
black holes as a function of time in an early galaxy with star formation
rate 10 \m\ yr$^{-1}$ for various values of the velocity parameter
$f_v = (v^2 + c_s^2)/c_s^2$ and $n_{ism} = 10^5$ cm$^{-3}$. Other 
parameters are set to fiducial values (see text). For larger values of 
$f_v$, the parameter $t_\infty$, the time for an initial black hole to 
grow to infinity, is large and the total energy has an asymptotic behavior 
(see text).} 
\end{figure}

\newpage

\begin{figure}[htp]
\centering
\includegraphics[totalheight=0.5\textheight]{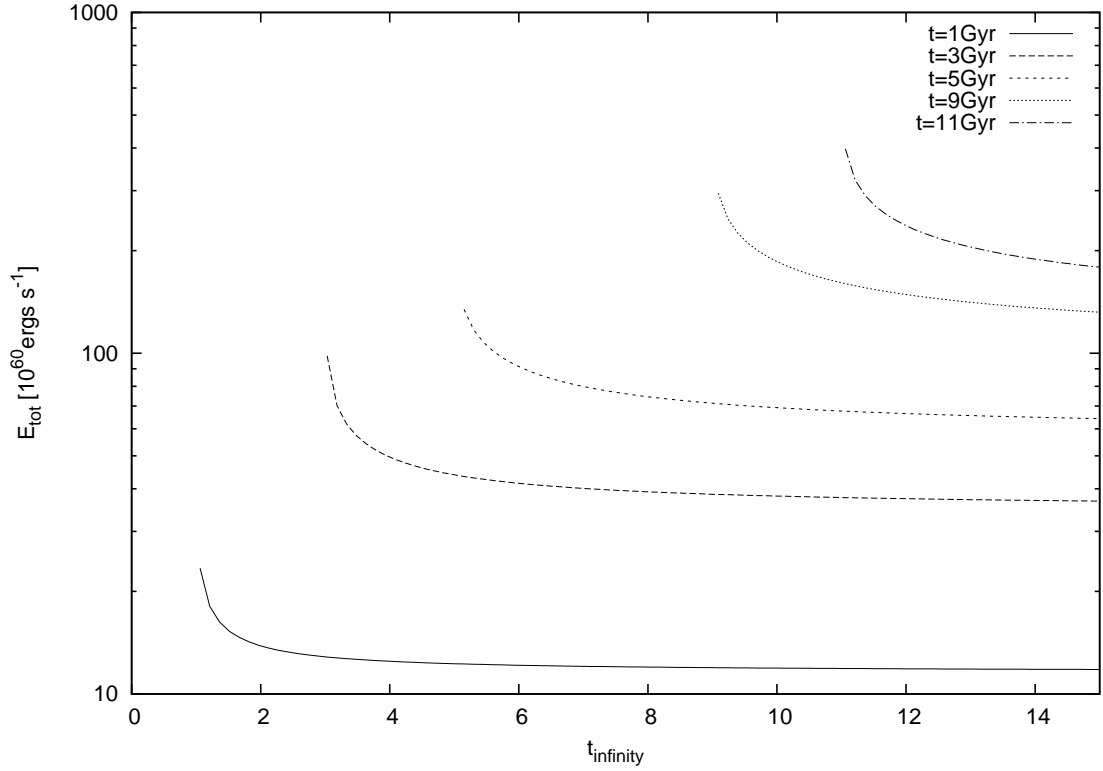}
\figcaption[Ebh_tinfinity.ps]{The total integrated energy of stellar-mass
black holes as a function of the parameter $t_\infty$, the time for 
an initial black hole to grow to infinity, for various fixed
epochs, t, in an early galaxy with star formation rate 10 \m\ yr$^{-1}$. 
The curves are truncated for t$_\infty$ approaching t from above due 
to graphics resolution.
}
\end{figure}

\end{document}